\documentstyle[fl eqn,twoside]{article}

\def\be{\begin{equation}}

\def\ee{\end{equation}}

\def\bi{\bibitem}

\begin{document}

\title{Canonical formulation of curvature squared action in the presence of lapse function}

\author{Abhik Kumar Sanyal$^{*}$, Subhra Debnath$^{\dag}$ and Soumendranath Ruz$^{\ddag}$}
\maketitle

\noindent

\begin{center}

$*,\dag$Dept. of Physics, Jangipur College, Murshidabad, India - 742213\\
$\ddag$Dept. of Physics, Satitara High School, Kandi, Murshidabad, India - 742170\\
\end{center}

\footnotetext[1]{
Electronic address:\\
\noindent $^{*}$sanyal\_ ak@yahoo.com\\
\noindent $^{\dag}$subhra\_ dbnth@yahoo.com\\
\noindent $^{\ddag}$ruzfromju@gmail.com\\}

\begin{abstract}

\noindent Lapse function appears as Lagrange multiplier in Einstein-Hilbert action and its variation leads to the $(^0_0)$ equation of Einstein, which corresponds to the Hamiltonian constraint equation. In higher order theory of gravity the situation is not that simple. Here, we take up the curvature squared $(R^2)$ action  being supplemented by an appropriate boundary term in the background of Robertson-Walker minisuperspace metric, and show how to identify the constraint equation and formulate the Hamiltonian without detailed constraint analysis. The action is finally expressed in the canonical form $A = \int(\dot h_{ij} \pi^{ij} + \dot K_{ij}\Pi^{ij} - N{\mathcal H})dt~ d^3 x$, where, the lapse function appears as Lagrange multiplier, once again. Canonical quantization yields Schr\"odinger like  equation, with nice features. To show that our result is not an artifact of having reduced the theory to a measure zero subset of its configuration space, the role of the lapse function as Lagrangian multiplier has also been investigated in Bianchi-I, Kantowski-Sachs and Bianchi-III minisuperspace metrics. Classical and semiclassical solutions have finally been presented.
\end{abstract}

\section{Introduction}
Explaining the cosmic evolution, taking only geometric terms in the action, has turned out to be an important issue presently,
since we do not have a scalar field at hand, as yet. Particularly, an action in the form, $A = \int [\alpha R + \beta R^2 + \gamma R^{-1}]\sqrt{-g} d^4 x$, apparently can challenge scalar field theories. The dominance of $R^{-1}$ term at the very late stage of evolution leads to effective negative pressure, sufficient to explain the SnIa data with an accelerating phase, the dominance of $R$ term in the middle, keeps the nucleosynthesis and the growth of perturbation necessary for structure formation, unchanged from Friedmann model, while the dominance of $R^2$ term in the early universe leads to inflation without invoking phase transition \cite {1}, \cite{2}. Modification of the Einstein-Hilbert action by including curvature squared terms ($R^2, R_{\mu\nu}R^{\mu\nu}, C_{\mu\nu\rho\lambda}C^{\mu\nu\rho\lambda}$), $C_{\mu\nu\rho\lambda}$ being the Weyl tensor, is also important in many other respect. It leads to a renormalizable theory of gravity \cite{3}, even while interacting with matter \cite{4} and is also asymptotically free \cite{5}. Unitarity of higher derivative quantum theory of gravity has also been established \cite{6}. Further, the Euclidean form of the Einstein-Hilbert action is not positive definite and therefore the functional integral corresponding to the ground state wave function of the universe diverges badly. A positive definite action that includes $R^2$-term, in the form \cite{ 7}, \cite{8}

\be S=-\frac{1}{4} \int~~d^4 X\sqrt{-g}~[A C_{ijkl}^2 +B (R-4\Lambda)^2],\ee

\noindent
leads to a convergent integral for the ground state wave function, and it reduces to Einstein-Hilbert action in the weak field limit. Additionally, canonical quantization of the above action leads to a Schr\"{o}dinger like equation, where an internal variable acts as the time parameter \cite{8}, \cite{9}, \cite{10}, \cite{11}. A string inspired theory of gravity \cite{12} and the $4$-dimensional Brane world effective action \cite{13} also contain such terms. In view of the above discussion, it turns out to be an important issue to include curvature squared term in the gravitational action and to study its quantum cosmological consequence, since it plays a dominating role only in the early universe. This requires canonical formulation of the theory.\\

\noindent
Canonical formulation of higher order theory of gravity requires a technique to reduce the Lagrangian to second order. This is usually performed considering (i) Lovelock action \cite{13a} (ii) scalar-tensor equivalence under conformal transformation \cite{13b} (iii) Ostrogradski's prescription \cite{14}. We shall follow Ostrogradski's prescription \cite{14} in which the phase space ($h_{i j}, K_{i j}, p^{i j}, P^{i j}$) is spanned by the first ($h_{ij}$) and the second ($K_{ij}$) fundamental forms together with their canonical momenta ($p^{ij}$) and ($P_{ij} = -2{\sqrt h}[2 A E^{ij} + B {(^4R)}h^{ij}]$) respectively, $E^{ij}$ being the electric part of the Weyl tensor (see \cite{14a} for more detail). Boulware \cite{15} modified the above prescription and proposed that the auxiliary variable may be chosen as $\pi_{i j}$, which is the momenta conjugate to the extrinsic curvature tensor $K_{i j}$. Following this prescription it has been observed that the above definition of $P_{i j}$ and that of canonical definition of the same, viz., $P_{i j} = \frac{\partial A}{\partial (\dot K^{i j})}$ are not at par, both in the isotropic and in some of the anisotropic models (see section 2 of \cite{16}). On the other hand, Horowitz \cite{8}, suggested that the auxiliary variable can be chosen as the `negative of the derivative of the action with respect to the highest derivative of the field variable present in the action.' In view of such a prescription, Horowitz \cite{8} obtained a Schr\"{o}dinger like equation instead of Wheeler-DeWitt type. It was Pollock \cite{17} who showed that, such prescription of auxiliary variable may be applied even in scalar tensor theory of gravity to yield totally different quantum dynamics. This prescription also allows to introduce auxiliary variable even in Einstein-Hilbert action, resulting in a quantum dynamics different from the Wheeler-DeWitt equation, which is not the correct description \cite{16}. Thus, the prescription given by Horowitz \cite{8} is somewhat misleading. The problem can be circumvented \cite{9,10,11}, \cite{16}, if Einstein-Hilbert action is supplemented with Gibbons-Hawking term and before attempting to introduce auxiliary variable, total derivative term is eliminated from the action, which gets cancelled with the boundary term. Once this is done, it is not possible to introduce auxiliary variable in Einstein-Hilbert action, even in the presence of a minimally or non-minimally coupled scalar field. Likewise, higher order theory of gravity should also be supplemented by appropriate boundary term and most importantly, some part of it must be eliminated prior to the introduction of auxiliary variable. In particular, for $R^2$ theory of gravity it is required to split the boundary term $\Sigma = 4\beta \int ~(^4R) K\sqrt {h} ~d^3 x$, into $\sigma_1 = 4\beta \int ~(^3R) K\sqrt {h} ~d^3 x$ and $\sigma_2 = 4\beta\int ~(^4R-^3R) K\sqrt {h} ~d^3 x$. In the process, $R^2$ theory of gravity becomes free from the trouble of boundary term and becomes as complete as Einstein-Hilbert action, when it is expressed as

\be A = \beta\int R^2\sqrt{-g}\;d^4 x  + \sigma_1 + \sigma_2.\ee
In the above, $K$ is the trace of the extrinsic curvature, $h$ is the determinant of the metric of the three space and $\beta$ is the coupling constant. From the above action, it is possible to eliminate only a single total derivative term which gets cancelled with $\sigma_1$. This follows automatically upon integration by parts, if one expresses the action in terms of the first fundamental form $h_{ij}$. It is then possible to introduce a unique auxiliary variable, following the suggestion of Horowitz \cite{8}. These facts had been elaborately discussed in \cite{16}. In \cite{16} it has been shown that upon quantization in terms of the basic variables ($h_{ij}$ and $K_{ij}$), an internal geometric parameter (the proper volume) acts as the time variable in the resulting Schr\"{o}dinger like equation. In the process, the effective Hamiltonian becomes hermitian and allows direct probability interpretation of the theory. Although, $\Sigma = \sigma_1 + \sigma_2$ has been found in view of Robertson-Walker minisuperspace metric, however it has also been tested successfully in a few other anisotropic minisuperspace line-elements, viz., Kantowski-Sachs, axially symmetric Bianchi-I and Bianchi-III \cite{16}. Thus the problem with boundary term has been alleviated at least for $R^2$ theory of gravity. It may be argued, why then one should not follow the easiest path through the scalar-tensor equivalence. The main reason is that it is a classical artifact. Remember, Hawking-Luttrell \cite{18} did not write the Wheeler-DeWitt equation for the scalar tensor equivalent action, due to the possibility that the conformal transformation may be singular. In the context of obtaining time dependent Schr\"{o}dinger like equations, we would like to mention that such equations have also been obtained by some authors \cite{18i, 18ii}, but from completely different perspective. In \cite{18i} an incoherent dust and in \cite{18ii} perfect fluid source act as the time parameters, instead of a geometric parameter. Hence, these are not our concern. \\

\noindent
Still there exists an unsolved issue. Upon (3 + 1) decomposition, lapse function ($N$) and shift vector ($N_i$) enter into the canonical form of the Einstein-Hilbert action as Lagrange multipliers, due to diffeomorphic invariance. Variation of lapse function yields Hamiltonian constraint equation, while variation of shift vector yields momentum constraint equations. Actions containing arbitrary curvature invariant term must show similar feature due to diffeomorphic invariance. However, this has not been attempted earlier, which we pose in the present work in the Robertson-Walker minisuperspace model, taking lapse function into account.\\

\noindent
The paper has been organized in the following manner. In section 2, we shall show how different auxiliary variable may be chosen in view of the proposals given by Boulware \cite{15} and Horowitz \cite{8} and discuss the associated problems. We also review Hawking-Luttrell's work \cite{18} in this connection. Finally, we show how the problem has been circumvented through the introduction of a unique auxiliary variable. Our main result appears in section 3, where taking Robertson-Walker mini-superspace in the presence of lapse function into account, we express the action containing curvature squared term in the canonical form $A = \int(\dot h_{ij} \pi^{ij} + \dot K_{ij}\Pi^{ij} - N{\mathcal H})dt~ d^3 x$, and the constrained Hamiltonian as $H_c = N \mathcal{H}$, where lapse acts as Lagrange multiplier. This issue has been tested in section 4 taking some of the anisotropic models, viz., Kantowski-Sachs, axially symmetric Bianchi-I and Bianchi-III into account. In section 5, the same has been continued for curvature squared action being supplemented by Einstein-Hilbert action. Classical solutions obtained by Starobinsky \cite{1} have been generated from the constrained Hamiltonian expressed in terms of the basic variables in section (6). Semiclasssical solution of the wave function corresponding to the quantum description of the theory under WKB approximation has been presented, which is peaked around the classical inflationary solution. In section 7, we have summarized our findings.

\section{A brief review of earlier attempts and aim of the present work}

\subsection{Earlier proposals to express $R^2$ action in canonical form}
Let us take scalar curvature squared action in the form,

\be A = \beta\int R^2\sqrt{-g} d^4 x. \ee

\noindent
Canonical formulation of such higher order theory of gravity requires a technique to reduce the Lagrangian to second order. One of the many routes (viz., Palatini formalism, lovelock formalism, scalar-tensor equivalence, loop quantum cosmology, use of auxiliary variable etc.) is to introduce auxiliary variable, performed following Ostrogradski's prescription \cite{14}. It suggests to fix a space like surface $\Sigma$ in a space-time $(M, g_{\mu\nu})$, which is asymptotically flat. Canonical variables are the three metric $h_{i j}$, a quantity $Q_{i j}$ (the auxiliary variable) related to the extrinsic curvature tensor and their conjugate momenta, $p^{i j}, P^{i j}$ respectively. Here, $Q_{i j}$ and $P^{i j}$ correspond to extra degree of freedom in higher derivative theory of gravity. The relations amongst $Q_{i j}, P^{i j}$, the space-time curvature $R$ and the extrinsic curvature $K_{i j}$ are

\be Q^{i j} = 2K^{i j},\;\;\;and\;\;\;P^{i j}= 8\beta\sqrt{h}h^{i j}~^4 R,\ee

\noindent
in the absence of electric part of the Weyl tensor\footnote {Boulware took the action in the form $A = -\frac{1}{4}B\int\sqrt{-g} d^4 x R^2$, and the definition of canonical momenta as, $P^{i j} = -2B\sqrt{h}h^{i j}~^4 R$}. Thus the phase space is spanned by $\{h_{i j}, Q_{i j} (K_{i j}), p^{i j}, P^{i j}\}$. Boulware \cite{15} modified the above prescription and proposed that the auxiliary variable may be chosen as $\pi_{i j}$, which is the momenta conjugate to the extrinsic curvature tensor $K_{i j}$, where, $\pi^{i j}=\int P^{i j}~d^3 x$, in the absence of the electric part of the Weyl tensor. Following this prescription it has been observed that the definition of $P_{i j}$ given in equation (4) and that of canonical definition of the same, viz., $P_{i j} = \frac{\partial A}{\partial (\dot K^{i j})}$ are not at par, both in the isotropic and in some of the anisotropic models (see section 2 of \cite{16}). Here we briefly illustration the situation in the isotropic case. The above action in the Robertson-Walker line element

\be ds^2 = - dt^2+a(t)^2\left[\frac{dr^2}{1-kr^2}+r^2 d\theta^2 + r^2\sin^2\theta d\phi^2\right],\ee

\noindent
takes the form
\be A = 36c\beta\int\left[\frac{\ddot a}{a} + \frac{\dot a^2}{a} + \frac{k}{a^2}\right]^2 a^3 dt,\ee

\noindent
where, $c = \int d^ 3 x$. Now, according to Boulware's  prescription \cite{15},

\be  Q^{ab} = 8\beta \int~^4 R  \sqrt{h} h^{ab} d^3 x = 8 c \beta R a.\ee

\noindent
On the other hand, the standard canonical definition of $Q^{ab}$ is,

\be Q^{ab} = \frac{\partial A}{\partial\dot K_{ab}}.\ee

\noindent
To calculate this, we remember, $K_{ab} = -a\dot a$, so let us make a change of variable $z = a^2$, which implies $\dot K_{ab} = -\frac{\ddot z}{2}$. The action then takes the form

\be A = 36 c \beta\int\left[\frac{\ddot z}{2 z} + \frac{k}{z}\right]^2 z^{\frac{3}{2}} dt.\ee

\noindent
Therefore,

\be \frac{\partial A}{\partial\dot K_{ab}} = -2\frac{\partial A}{\partial \ddot z} = -12 c \beta R a.\ee
Thus the two definitions of canonical momenta (4) and (7) yield different results, viz., (8) and (10) respectively and as such do not match. Just changing the factor 8 by -12 in Boulware's  prescription \cite{15} does not solve the problem, since it reappears in anisotropic model \cite{16}.\\

\noindent
Horowitz \cite{8} prescribed to choose the auxiliary variable as the `negative of the derivative of the action with respect to the highest derivative of the field variable present in the action.' Eventually, asymptotic flatness condition required by the prescription given by  Ostrogadski is not required any further. Now according to Horowitz \cite{8}, the auxiliary variable is either $-12 c\beta R a^2$, taking the scale factor $a$ as the basic variable or $-6 c\beta R a$, taking $z = a^2 = h_{ij}$ as the basic variable and the question arises which one should be treated as basic variable. This prescription was applied by Pollock taking $a$ as the basic variable which we shall discuss in subsection (2.2) \cite{17,18a}. \\

\noindent
Hawking and Luttrell \cite{18} took Euclidean action containing $(R + \beta R^2)$ term in the conformal form of the Robertson-Walker line element

\be ds^2= \frac{4a^2}{3\pi} [d\eta^2 + d\Omega^2],\ee

\noindent
for which Ricci scalar reads
\be R = \frac{9\pi}{2}\left(\frac{1}{a^2} -\frac{a''}{a^3}\right), \ee

\noindent
while the action is

\be \hat{I} = - \int d\eta\left[\frac{1}{a^2} -\frac{a''}{a^3} + \frac{9}{2}\pi\beta\left(\frac{1}{a^2} -\frac{a''}{a^3}\right)^2\right]a^4,\ee

\noindent
and choose an auxiliary variable in the form $Q = a(1+2\beta R)$ to reduce the fourth order equations to second order treating $a$ and $Q$ to be independent variables \footnote{Note that the surface term contains $2\int d^3 x \sqrt{h}[K(1+2\beta R)]$}. The idea behind such a choice is that, under the conformal transformation $\tilde{g}_{\mu\nu} = (1+2\beta R) g_{\mu\nu}$, scalar-tensor equivalence is established. This means, Wheeler-DeWitt equation obtained under such a choice of auxiliary variable appears to resemble with that for ordinary Einstein gravity coupled to a massive scalar field. Thus they attempted to mimic the effect of a massive scalar field in the gravitational Lagrangian. Such technique for reducing higher order terms was also followed by Mazzitelli \cite{18b} for the purpose of renormalization. Thus, $a$ and $Q$ are treated as basic variables in this formalism, which may appear to be dependent variables at first sight. Nevertheless, Kaspar \cite{18b1} made a rigorous constraint analysis to explore that indeed the two formalisms viz., the one given by Buchbinder and Lyachovich \cite{18b2} and that followed by Hawking-Luttrell \cite{18} are essentially the same, which implies $a$ and $Q$ may be treated as independent variables. One can also retrieve Hawking-Luttrell's choice of auxiliary variable \cite{18} following Horowitz's \cite{8} prescription as,

\be \frac{\partial \hat{I}}{\partial a''} = a(1+2\beta R).\ee

\noindent
If we throw away the factor $\frac{4}{3\pi}$ from the metric (11), the result is $\frac{\partial \hat{I}}{\partial a''} = 6ca(1+2\beta R)$. Further, if one translates the metric to non-conformal form, the result certainly mimics Horowitz prescription (with $a$ as basic variable). All these calculations are performed to understand that for canonical formulation of higher order theory of gravity there is no unique choice of auxiliary variable. The question is how far such different choice of auxiliary variables, affect classical and quantum dynamics? In this context we quote a sentence from Kaspar \cite{18b2}, viz., ``The choice of momentum operators is of fundamental importance if one is thinking
of observable''. The choice of momentum operator clearly depends upon the choice of auxiliary variable and so a judicious choice of the same is required. It is also important to note that the auxiliary variable chosen by Hawking and Luttrell \cite{18} and the prescription given by Horowitz \cite{8} do not keep Einstein-Hilbert sector apart from $R^2$ sector. That is to say, they suggest to introduce auxiliary variable in the linear sector also. This is the reason why, such prescription does not yield Einstein theory in the weak field (small curvature) limit. This fact was first noticed by Pollock \cite{17}.\\

\subsection{Problem associated with such proposals}
\noindent
Taking a toy model into account, Pollock \cite{17} showed that, quantum version of the following action
\be A = \int\sqrt{-g} d^4 x\left(\frac{1}{2}\epsilon\phi^2 R - \frac{1}{4}\lambda \phi^4\right),\ee

\noindent
using the prescription given by Horowitz \cite{8}, leads to the following schr\"{o}dinger-like equation (where, $x = \alpha'$ and $\alpha = t$, in the conformal form of Robertson-Walker metric),
\be ix \frac{\partial\Psi}{\partial t} = \frac{1}{m} \frac{\partial^2\Psi}{\partial x^2} + i\frac{\partial \Psi}{\partial x} + ix^{(2-p)}\frac{\partial (x^p\Psi)}{\partial x},\ee

\noindent
in the unit $\hbar = 1$, where, $p$ is the factor ordering index. Equation (16) yields time independent solution in the form

\be \Psi(t, x) = \exp\left[{-im\left(x + \frac{1}{3}x^3\right)}\right].\ee

\noindent
Equation ()17 is not the correct quantum description of the model, since Einstein-Hilbert action must lead to Wheeler-DeWitt equation, instead. It may be easily shown that one can introduce auxiliary variable even in Einstein-Hilbert action. For the purpose let us choose the action in the form

\be A = \int\sqrt{-g}d^4 x\left[\frac{R}{16\pi G} - \Lambda\right] + \frac{1}{8\pi G} \int\sqrt{h} K d^3 x,\ee

\noindent
which in the isotropic metric (5) takes the form (absorbing $\kappa = \frac{8\pi G}{3}$ in the action)

\be A = c\int[ a^2 \ddot a + a\dot a^2 + k a - \kappa\Lambda a^3 ]dt + \frac{1}{3} \int\sqrt{h} K d^3 x. \ee

\noindent
The auxiliary variable,
\be Q = - \frac{\partial A}{\partial \ddot a} = -ca^2,\ee

\noindent
now may be introduced into the action as,
\be A =\int\left[-Q\ddot a - (\dot a^2 + k)\frac{Q}{a} - c\kappa\Lambda a^3\right]dt - c a^2 \dot a.\ee

\noindent
Under integration by parts, the total derivative term gets cancelled with the Gibbons-Hawking boundary term \cite{18c}, and the action is automatically cast in the canonical form as,

\be A = \int\left[\dot Q \dot a - (\dot a^2 + k)\frac{Q}{a} - c\kappa\Lambda a^3\right]dt.\ee

\noindent
The Hamiltonian is

\be H = p_a p_Q + \frac{Q}{a}p_{Q}^2 + k\frac{Q}{a} + c\kappa\Lambda a^3,\ee

\noindent
To express the above Hamiltonian in terms of the basic variables, let us make a change of variable, $\dot a = x$. It is now apparent that since, $Q = - \frac{\partial A}{\partial \dot x}$, so $Q$ should be replaced by $-p_x$, while, since $p_Q = \dot a$, so $p_Q$ should be replaced by $x$. Thus the above Hamiltonian takes the form

\be H = x p_a - \left(\frac{x^2}{a} + \frac{k}{a}\right)p_x + c\kappa\Lambda a^3 = 0.\ee

\noindent
Under quantization it yields,

\be i\hbar a \frac{\partial \Psi}{\partial a} = i\hbar (x^2+k)\frac{\partial \Psi}{\partial x} + c\kappa\Lambda a^4\Psi.\ee

\noindent
Thus instead of Wheeler-DeWitt equation, a totally different equation emerges, which gives completely wrong quantum dynamics. It is wrong because even at the classical level, whatever matter be introduced in the action (18), $Q$ variation equation under such technique, always yields $R = 0$. This clearly reveals that under any circumstances auxiliary variable should not be introduced in Einstein-Hilbert sector. Much convincing argument may be given in terms of the action containing Gauss-Bonnet-Dilatonic coupling in 4-dimension. Despite the fact that Gauss-Bonnet term is constructed from higher order curvature invariant terms, canonization does not require auxiliary variable \cite{18d}. This is because, it is first integrated by parts to remove total derivative term and one is left with standard form of the point Lagrangian $L = L(a, \phi, \dot a, \dot\phi.)$. From the above discussion it should be clear that Horowitz's prescription \cite{8} enforces to introduce auxiliary variable in linear gravity and also in Gauss-Bonnet gravity, leading to wrong quantum dynamics. In this connection, Horowitz's  proposal \cite{8} was modified by Sanyal and Modak and Sanyal \cite{9, 10, 11}, to circumvent the problem.

\subsection{Resolving the issue}

Proposal given by Sanyal and Modak and Sanyal \cite{9, 10, 11} states that auxiliary variable should be introduced only after the removal of all available total derivative terms from the action. In the process Einstein-Hilbert action becomes free from second derivative term and the total derivative term gets cancelled with Gibbons-Hawking boundary term \cite{18c}. Thus the problem with Einstein-Hilbert action gets resolved, there is no scope to introduce auxiliary variable. However, this is not all, since such prescription creates problem in $R^2$ action. This problem was finally resolved by Sanyal \cite{16}. Here we briefly illustrate the problem and its resolution underneath. $R^2$ action must be supplemented by a boundary term $\Sigma = 4\beta\int~^4R\sqrt{h}K d^3x$ which appears upon delta variation of the action. So the complete action should be written as

\be A = \beta\int~R^2 \sqrt{g} d^4x + \Sigma,\ee

\noindent
which in the above Robertson-Walker line element (5) reads,

\be A = M \int\left[a\ddot a^2 + \frac{(\dot a^2 + k)^2}{a} + 2\dot a^2\ddot a + 2k\ddot a\right] dt + \Sigma,\ee

\noindent
where, $M = 36\beta c$. Under integration by parts, it is expressed as,

\be A = M \int\left[a\ddot a^2 + \frac{(\dot a^2 + k)^2}{a}\right]dt + 2M\left[\frac{1}{3}\dot a^3 + ka\right] +\Sigma.\ee

\noindent
Now one can introduce the auxiliary variable (negative of the action with respect to the highest derivative present as proposed by Horowitz to introduce auxiliary variable is not important, since it can be taken care of later)

\be Q = \frac{\partial A}{\partial \ddot a} = 2Ma\ddot a,\ee

\noindent
in the above action as

\be A = \int\left[Q\ddot a - \frac{Q^2}{4M a} + \frac{(\dot a^2 + k)^2}{a}\right]dt + 2M\left[\frac{1}{3}\dot a^3 + ka\right] +\Sigma.\ee

\noindent
Again upon integration by parts, the action is expressed in the following canonical (with non-vanishing Hessian determinant) form,

\be A = \int\left[-\dot Q\dot a - \frac{Q^2}{4M a} + \frac{(\dot a^2 + k)^2}{a}\right]dt + 2M\left[\frac{Q\dot a}{2M}+\frac{1}{3}\dot a^3 + ka\right] +\Sigma.\ee

\noindent
Although the above action yields correct classical field equations, however the total derivative term thus obtained $2M[a\dot a\ddot a + \frac{1}{3}\dot a^3 + ka]$ does not cancel the boundary term $\Sigma = -2M[a\dot a\ddot a + \dot a^3 + ka]$. This problem was cured by Sanyal \cite{16} by taking the first fundamental form $h_{ij} = z = a^2$ as the basic variable, instead of the scale factor $a$. Remember, this was originally suggested by Ostrogradski \cite{14} and Boulware \cite{15}, but the associated problem was not encountered by earlier authors, since auxiliary variable was introduced a-priori, and upon integration by parts, total derivative term gets cancelled with the boundary term. Now taking $z$ as the basic variable, the action becomes

\be A = M \int\left[\frac{\ddot z^2}{4\sqrt z} + \frac{k^2}{\sqrt z} + \frac{k\ddot z}{\sqrt z}\right]dt + \Sigma.\ee

\noindent
Upon integration by parts, it yields

\be A = M \int\left[\frac{\ddot z^2}{4\sqrt z} + \frac{k^2}{\sqrt z} + \frac{k\dot z^2}{2 z^{\frac{3}{2}}}\right]dt + M\frac{k\dot z}{\sqrt z} +\Sigma.\ee

\noindent
It is interesting to note that $\Sigma$ may be split up in two terms as $\Sigma = \sigma_1 + \sigma_2$, where, $\sigma_1 = 4\beta\int ~^3RK \sqrt {h} d^3 x$, which gets cancelled with the total derivative term $M\frac{k\dot z}{\sqrt z}$, obtained above and $\sigma_2 = 4\beta\int(^4R - ^3RK) \sqrt {h} d^3 x$. The auxiliary variable

\be Q = \frac{\partial A}{\partial \ddot z} = \frac{M\ddot z}{2\sqrt{z}}, \ee

\noindent
(different from $R a$) may now be introduced to express the above action in the following canonical form

\be A = \int\left[Q\ddot z - \frac{\sqrt {z}Q^2}{M} + \left(\frac{k^2}{\sqrt z} + \frac{k\dot z^2}{2 z^{\frac{3}{2}}}\right)\right]dt + \sigma_2.\ee

\noindent
Upon integration by parts, the total derivative term gets cancelled with the boundary term $\sigma_2$ and the final canonical action is

\be A = \int\left[-\dot Q\dot z - \frac{\sqrt {z}Q^2}{M} + \left(\frac{k^2}{\sqrt z} + \frac{k\dot z^2}{2 z^{\frac{3}{2}}}\right)\right]dt.\ee

\noindent
Quantization of the above action and the one with Einstein-Hilbert term in addition, were studied by Sanyal \cite{16}. So, in a nutshell, the scheme is to express the action containing $R^2$ term, in terms of the basic variables $h_{ij}$ and total derivative term should be eliminated, which gets cancelled with $\sigma_1$. Auxiliary variable should be introduced thereafter as suggested by Horowitz \cite{8} judiciously, to express the action in the canonical form (negative sign in the choice of auxiliary variable, as already mentioned, is of no importance, since it may be taken care of, during quantization). Upon integration by parts, $\sigma_2$ gets cancelled, and the final form of the action gives correct classical field equations and Schr\"{o}dinger like equation emerges upon quantization, with an hermitian effective Hamiltonian and standard (quantum mechanical) probability interpretation.

\subsection{What is still left: aim of present work}

\noindent
Still, there remains an unsolved important issue, which, as explained underneath, is presently our concern. The space-time metric $ds^2 = g_{\mu\nu} dx^{\mu}dx^{\nu}$ under $(3 + 1)$
decomposition can be expressed as,

\be ds^2= -\left(N^2-N_i N^i\right)dt^2+ 2N_i dx^i dt + h_{ij}dx^i dx^j, \ee
where, $N$ and $N_i$ are the lapse function and shift vector respectively. In view of such decomposition, Einstein-Hilbert action when supplemented by Gibbons-Hawking surface term \cite{18c},

\be A = \int\frac{1}{16\pi G}(R-2\Lambda)\sqrt{-g}\;d^4 x  + \frac{1}{8\pi G} \int\sqrt h ~ K ~ d^3 x, \ee
leads to a canonical action in terms of the basic variables $h_{ij}$ and its canonical conjugate momenta $\pi_{ij}$, in the form,

\be A =\int\left[\dot h_{ij}\pi^{ij} -H_c - H_{ci}\right]d^3 x dt = \int\left[\dot h_{ij}\pi^{ij} -N{\mathcal H} - N^i {\mathcal H}_{i}\right]d^3 x dt,\ee
where, due to diffeomorphic invariance, the lapse function $N$ and the shift vector $N^i$ appear as Lagrange multipliers. Variation of the action (39) with respect to the shift vector gives the super momentum constraint,

\be {\mathcal{H}}_{i} = 2D_{j}\pi{_{i}}^j = 0,\ee
and variation with respect to the lapse function gives the super Hamiltonian constraint,

\be {\mathcal H} = (16\pi G)~ G_{ijkl}\pi^{ij}\pi^{kl} - \frac{1}{16\pi G}\sqrt h~(^3 R -2\Lambda) = 0,\ee
where, the metric on the superspace is expressed as,

\be G_{ijkl} = \frac{1}{2\sqrt h}(h_{ik}h_{jl}+h_{il}h_{jk}-h_{ij}h_{kl}).\ee

\noindent
Canonical quantization is then straight forward, which gives the so called Wheeler-DeWitt equation. Thus to show that the action (2) corresponding to $R^2$ gravity, is as complete as Einstein-Hilbert action (38), it should also be expressed in the canonical form (39) taking into account the lapse function and the shift vector. Such a construction has not been found possible considering the whole superspace, and it has not so far been attempted in minisuperspace models too. In the following section, we attempt this issue in the Robertson-Walker minisuperspace model, which accommodates lapse function $N(t)$ only. We follow the prescription \cite{16} and in the process show that the lapse function $N$ here again appears as a Lagrange multiplier and the action is expressible in the canonical form in terms of the basic variables $h_{i j}, K_{i j}$, as suggested in Boulware \cite{15}. In section 4, we take up anisotropic Bianchi-I, Kantowski-Sachs and Bianchi-III metrics to exemplify that our construction for $R^2$ theory of gravity is indeed correct. In section 5, we extend our work to include Einstein-Hilbert action in addition to curvature squared term. In section (6) classical solutions of the field equations are presented, which are at par with those given by Starobinskii \cite{1}. Semiclassical solution under WKB approximation has also been presented in the same section, which is peaked around the classical inflationary solution. Finally we conclude in section (7).

\section{Canonical formulation of (scalar) curvature squared action}
In our earlier work \cite{16}, we have shown that under appropriate choice of auxiliary variable the boundary term corresponding to $R^2$ action may be represented as a couple of total derivative terms and quantization leads to Schr\"odinger like equation. Here, in the background of Robertson-Walker minisuperspace, we intend to show that in the process of canonical formulation of $R^2$ gravity, the lapse function $N(t)$ acts as Lagrange multiplier, while the action may be expressed in canonical form in terms of the basic variables $h_{i j}$ and $K_{i j}$. The variation of the action gives the Hamiltonian constraint equation as in the case of Einstein-Hilbert action. Let us consider a general scale invariant action,

\be A = \int \left[\alpha C_{\mu\nu\rho\lambda}C^{\mu\nu\rho\lambda} + \beta R^2\right]\sqrt{-g} d^4 x.\ee
Since the Weyl tensor vanishes in the Robertson-Walker minisuperspace metric,

\be ds^2 = -N(t)^2dt^2+a(t)^2\left[\frac{dr^2}{1-kr^2}+r^2 d\theta^2 + r^2\sin^2\theta d\phi^2\right],\ee
hence the above action, in the presence of a cosmological constant $\Lambda$ and being supplemented by appropriate boundary term, as in reference [19], reduces to

\be A = \beta\int\sqrt{-g}\;d^4 x\left[R^2
-\frac{2\Lambda}{\beta}\right] + \sigma_1 + \sigma_2,\ee \noindent
The Ricci scalar, $R = \frac{6}{N^2}\left(\frac{\ddot a}{a}+\frac{\dot a^2}{a^2}+N^2\frac{k}{a^2}-\frac{\dot N\dot a}{N a}\right)$, under the choice $h_{ij} = a^2 = z$ takes the form,

\be R = \frac{6}{N^2}\left[\frac{\ddot z}{2z}+N^2\frac{k}{z}-\frac{\dot z\dot N}{2zN}\right],\ee
and the above action now reads,

\be A = c\beta\int\left[\frac{9\ddot z^2}{N^3\sqrt z}+36k\left(\frac{\ddot z}{N\sqrt z}-\frac{\dot N\dot z}{N^2\sqrt z}\right)-\frac{18\dot N\dot z\ddot z}{N^4\sqrt z}+\frac{9\dot N^2\dot z^2}{N^5 \sqrt z}+\frac{36Nk^2}{\sqrt z}-\frac{2N\Lambda z^{\frac{3}{2}}}{\beta}\right]dt+\sigma_1+\sigma_2,\ee
where, as mentioned in the introduction,

\be \Sigma = \sigma_1 + \sigma_2 = 4\beta\int\sqrt h ~K ~^3R~ d^3 x + 4\beta\int\sqrt h ~K (^4R-^3R)~ d^3 x = 4\beta\int{^4R}~\sqrt h ~K ~ d^3 x\ee
and the constant $c$ is the volume of the three space. Under integration by parts the first bracketed terms in the above action yield a counter term, that gets cancelled with  $\sigma_1$ and we are left with,

\be A = B\int\left[\frac{9\ddot z^2}{N^3\sqrt z}-\frac{18\dot N\dot z\ddot z}{N^4\sqrt z}+\frac{9\dot N^2\dot z^2}{N^5 \sqrt z}+\frac{18k\dot z^2}{Nz^{\frac{3}{2}}}+\frac{36Nk^2}{\sqrt z}-\frac{2N\Lambda z^{\frac{3}{2}}}{\beta}\right]dt+\sigma_2,\ee
where, we have chosen $B = c\beta$. At this stage we introduce the auxiliary variable,

\be Q = \frac{\partial A}{\partial \ddot z} = 18B\left[\frac{\ddot z}{N^3 \sqrt z} - \frac{\dot N \dot z}{N^4 \sqrt z}\right]\ee
and express the action in the canonical form as,

\be A = B\int\left[\frac{Q\ddot z}{B} - \frac{\dot N\dot z}{B N} Q - \frac{N^3 \sqrt z}{36 B^2} Q^2 + \frac{18k\dot z^2}{Nz^{\frac{3}{2}}}+\frac{36Nk^2}{\sqrt z}- \frac{2N\Lambda z^{\frac{3}{2}}}{\beta}\right]dt + \sigma_2.\ee
Shortly, we shall prove our claim that the above action (51) indeed may be expressed in the canonical form as in (39), in view of the basic variables $\{h_{i j}, K_{i j}, p^{i j}, P^{ij}\}$. Now the first term in (51) is integrated by parts and the total derivative term gets cancelled with $\sigma_2$, and we are finally left with,

\be A = \int\left[-\dot Q\dot z - \frac{\dot N}{N}\dot z Q + \frac{18B k\dot z^2}{Nz^{\frac{3}{2}}}- \frac{N^3 \sqrt z}{36 B} Q^2 +B\left(\frac{36 Nk^2}{\sqrt z}-\frac{2\Lambda}{\beta} N z^{\frac{3}{2}}\right)\right]dt.\ee
The canonical momenta are,

\be p_z = -\dot Q - \frac{\dot N}{N}Q+36 Bk\frac{\dot z}{N z^{\frac{3}{2}}},\;\;p_Q = -\dot z, \;\;p_N = -\dot z\frac{Q}{N}.\ee
The $Q$ variation equation gives back the definition of $Q$ given in (50), while the $z$ variation equation is,

\be \ddot Q + \frac{\ddot N}{N}Q + \frac{\dot N}{N}\dot Q - \frac{\dot N^2}{N^2}Q - 36B k \left(\frac{\ddot z}{N z^{\frac{3}{2}}} - \frac{\dot N \dot z}{N^2 z^{\frac{3}{2}}} -\frac{3\dot z^2}{4Nz^{\frac{5}{2}}}\right) -\frac{N^3 Q^2}{72 B\sqrt z} - 18 B k^2\frac{N}{z^{\frac{3}{2}}} + \frac{3B}{\beta}N\Lambda\sqrt z = 0 \ee
and the $N$ variation equation is

\be -\frac{\ddot z Q}{N} - \frac{\dot z\dot Q}{N} + 18Bk\frac{\dot z^2}{N^2 z^{\frac{3}{2}}}+ \frac{N^2 \sqrt z}{12B}Q^2  - B\left(\frac{36 k^2}{\sqrt z}-\frac{2\Lambda}{\beta} z^{\frac{3}{2}}\right) = 0.\ee

\noindent
At this end we need a little discussion which we itemize underneath.
\begin{itemize}

\item In view of the definition of momenta given in (53), it is clear that neither $Q$ nor $N$ is invertible, which signals the presence of a constraint in the theory. This is also apparent from the fact that the Hessian determinant vanishes, i.e., ${\mathbf{H}} = |\sum_{i,j} \frac{\partial^2 L}{\partial \dot q_i\partial\dot q_j}| = 0$. Apparently, $Q$ and $N$ are having the same status in the action. Nevertheless, we know that $Q$ is only an auxiliary variable and it has been introduced in the action keeping its canonical form intact. As a result $Q$ variation equation gives back the definition (50) of $Q$ and must not give any dynamics. Further, in view of diffeomorphic invariance it is known that $N$ must act as a Lagrange multiplier.

\item Now, the second signal for the presence of constraint is that a particular variable is non-dynamical, i.e., none of the field equations should contain second derivative of that variable. But here we observe that second derivative of $N$ appears in equation (54). This is definitely confusing.

\item Finally, the third signal for the presence of a constraint is that one of the field equations must not be dynamical, i.e., it must not contain second derivative term. However, both the equations (54) and (55) contain second derivative terms. Thus, action (52) contains a constraint, but the constraint equation is hidden.

\end{itemize}
The reason is, unlike the case of Einstein-Hilbert action, the action (52) contains first derivative of $N$. In fact, one can choose, a variable $q = N Q$ to get rid of $\dot N$ term from the action (52). In the process, $N$ acts as Lagrange multiplier and $\ddot N$ term disappears from the $z$ variation equation (54) while, $\ddot z$ term disappears from the $N$ variation equation (55). As a result, equation (55) being free from second derivative term, stands as the constraint equation (see appendix). Instead, one can also remove $\ddot z$ term from equation (55) in view of the definition of $Q$ given in (50). Thus equation (55) takes the following form, viz.,

\be -\frac{\dot z \dot Q}{N} - \frac{\dot N\dot z Q}{N^2} + 18Bk\frac{\dot z^2}{N^2 z^{\frac{3}{2}}} + \frac{N^2 \sqrt z}{36B}Q^2  - B\left(\frac{36 k^2}{\sqrt z} - \frac{2\Lambda}{\beta} z^{\frac{3}{2}}\right)= 0.\ee

\noindent
This is the equation we were in search of, which does not contain second derivative term and hence is a constraint of the system under consideration. It can be easily verified that this is the Hamiltonian of the system in disguise,

\be H_c = N\left[-\frac{\dot z \dot Q}{N} - \frac{\dot N\dot z Q}{N^2} + 18Bk\frac{\dot z^2}{N^2 z^{\frac{3}{2}}} + \frac{N^2 \sqrt z}{36B}Q^2  - B\left(\frac{36 k^2}{\sqrt z} - \frac{2\Lambda}{\beta} z^{\frac{3}{2}}\right)\right],\ee

\noindent
which is constrained to vanish in view of equation (56). So, as in the case of Einstein-Hilbert action, here too the Hamiltonian can be obtained under the variation of the lapse function, but for that, one has to utilize the definition of auxiliary variable in addition. Now, the next question is, how to express the Hamiltonian in terms of the phase-space variables? This usually requires detailed constraint analysis \cite{19}, where, the constraint in the configuration space variable is equation (56) and that in phase space variable is $Qp_Q - Np_N = 0$, as is observed in view of the canonical momenta (53). However, we show that even without going into the details of constraint analysis, the Hamiltonian in terms of the phase space variables may be obtained in a straightforward manner. This is possible because $N$ acts only as a Lagrange multiplier. The definitions of canonical momenta (53) yield,

\be p_Q p_z = \dot z\dot Q + \frac{\dot N}{N}\dot z Q - 36Bk\frac{\dot z^2}{N z^{\frac{3}{2}}}.\ee So,

\be -\dot z \dot Q - \frac{\dot N\dot z Q}{N} + 18Bk\frac{\dot z^2}{N z^{\frac{3}{2}}} = - p_Q p_z -18 Bk\frac{\dot z^2}{N z^{\frac{3}{2}}} = - p_Q p_z -  \frac{18Bk}{N z^{\frac{3}{2}}}p_{Q}^2,\ee

\noindent
where, we have replaced $\dot z$ by $p_Q$, instead of $p_{N}$, since $N$ acts as Lagrange multiplier and so the Hamiltonian must not contain $p_{N}$. Thus the Hamiltonian constraint equation in terms of the phase space variables is obtained as,

\be H_c = -p_Q p_z -  \frac{18 Bk}{N z^{\frac{3}{2}}}p_{Q}^2 + \frac{N^3 \sqrt z}{36B}Q^2  -BN\left(\frac{36 k^2}{\sqrt z}-\frac{2\Lambda}{\beta} z^{\frac{3}{2}}\right) = 0.\ee

\noindent
This is not the end of the story. To express the curvature squared action in the canonical form (39) as in the case of Einstein-Hilbert action, we need to express $H_c$ as $H_c = N{\mathcal H}$. But, first of all it is required to express the Hamiltonian in terms of the basic variables (instead of auxiliary variable) spanned by, \{$h_{ij}, \pi_{ij}, K_{ij}, \Pi_{ij}$\} as par Ostrogradski's prescription \cite{14}, which essentially are \{$z, p_z, \dot z, p_{\dot z}$\} in the Robertson-Walker minisuperspace metric. For this purpose and to avoid confusion, the standard choice is $\dot z = -2K_{ij} = x$, in the absence of the lapse function ( see \cite{8}, \cite{9}, \cite{10}, \cite{11}, \cite{16}). Thus, our extended phase space is spanned by \{$z, p_z, x, p_{x}$\}. But here in the presence of lapse function, $K_{ij} = \frac{\dot h_{ij}}{2 N}$, so  we choose,

\be x = \frac{\dot z}{N}.\ee

\noindent
Note that in the process again, $x = -2K_{ij}$. Hence,

\be Q = \frac{\partial A}{\partial \ddot z} = \frac{\partial A}{\partial \dot x}\frac{d\dot x}{d\ddot{z}} = \frac{p_x}{N},\;\;and,\;\;p_{Q} = -\dot z = -N x.\ee

\noindent
Therefore we need to replace $Q$ by $\frac{p_x}{N}$ and $p_{Q}$ by $-Nx$ in the above Hamiltonian. Thus, finally we are able to write,

\be H_c = N\left(xp_z + \frac{\sqrt z}{36B}p_{x}^2 - \frac{18 Bk}{z^{\frac{3}{2}}}x^2  - 36B\frac{k^2}{\sqrt z}+\frac{2B\Lambda}{\beta} z^{\frac{3}{2}}\right)= N{\mathcal{H}} = 0.\ee

\noindent
It is now straightforward to express the action (51) as [since $\dot z = N x$, therefore, we substitute $\ddot z = N\dot x + \dot N x$, in the first term of (51), $\dot z = Nx, \; Q = \frac{p_x}{N}$, in the second and third terms, $x = \frac{\dot z}{N}$, in the fifth and $p_{x}^2 = N^2 Q^2$ in the sixth],

\be A = \int\left(\dot z p_z + \dot x p_x - N\mathcal{H}\right)dt~ d^3 x,\ee

\noindent
and we have achieved our goal of expressing the curvature squared action in the canonical form with respect to the basic variables. We can also anticipate a general form of the above canonical action. Remember, $x = \frac{\dot z}{N} = 2\frac{a\dot a}{N} = -2 K_{ij}$, and so $\dot x = -2\dot K_{ij}$, where, $K_{ij}$ is the extrinsic curvature tensor. Again, $p_x = -\frac{1}{2}\Pi_{ij}$, $\Pi_{ij}$ being the momentum canonically conjugate to $K_{ij}$. Hence we may write the above form of canonical action as,

\be A = \int\left(\dot h_{ij} \pi^{ij} + \dot K_{ij}\Pi^{ij} - N\mathcal{H}\right)dt~ d^3 x,\ee

\noindent
where, $\pi^{ij}$ is the momenta canonical to $h_{ij}$. This result that curvature squared action can be expressed in a general canonical form is of course new and exciting, even though it has been done in a minisuperspace metric. The most important point is to note that, we have chosen the auxiliary variable as was suggested by Horowitz \cite{8}, but not before a total derivative term present in the action has been taken care of. In the process, we have finally been able to express the action in the canonical form in terms of the basic variables as par Ostrogradski's prescription. Now, the canonical quantization of the Hamiltonian constraint equation (63) is straightforward, which yields,

\be \frac{i\hbar}{\sqrt z}\frac{\partial \Psi}{\partial z} = -\frac{\hbar^2}{36B x}\left(\frac{\partial^2}{\partial x^2} + \frac{n}{x}\frac{\partial}{\partial x}\right)\Psi -18 Bk \left(\frac{x}{z^2}+\frac{2k}{z x}\right)\Psi + \frac{2B\Lambda}{\beta} \frac{z}{x}\Psi,\ee

\noindent
where, $n$, the factor ordering index, removes some (but not all) factor ordering ambiguities. Again under a further change of variable, the above equation takes the look of the Schr{\"o}dinger equation, viz.,

\be i\hbar\frac{\partial \Psi}{\partial \alpha} = -\frac{\hbar^2}{54B}\left(\frac{1}{x}\frac{\partial^2}{\partial x^2} + \frac{n}{x^2}\frac{\partial}{\partial x}\right)\Psi - B \left[\frac{12kx}{\alpha^{\frac{4}{3}}}+\frac{24 k^2}{x\alpha^{\frac{2}{3}}} - \frac{4\Lambda}{3\beta x}\alpha^{\frac{2}{3}}\right]\Psi = \hat H_e\Psi,\ee

\noindent
where, $\alpha  = z^{\frac{3}{2}} = a^3$. Hence, the proper volume plays the role of internal time parameter. Note that the effective Hamiltonian

\be \hat H_e(x, \alpha) = -\frac{\hbar^2}{54B}\left(\frac{1}{x}\frac{\partial^2}{\partial x^2} + \frac{n}{x^2}\frac{\partial}{\partial x}\right)+V_e(x,\alpha),\ee

\noindent
is hermitian, where the effective potential $V_e$, given by,

\be V_e(x, \alpha) = -B \left[\frac{12kx}{\alpha^{\frac{4}{3}}}+\frac{24 k^2}{x\alpha^{\frac{2}{3}}} - \frac{4\Lambda}{3\beta x}\alpha^{\frac{2}{3}}\right],\ee

\noindent
is a function of both the so called time variable $\alpha$ and $x$. The hermiticity of the effective Hamiltonian allows one to write the continuity equation for $n = -1$, as,

\[\frac{\partial\rho}{\partial\alpha} + \nabla . {\bf{J}} = 0,\]

\noindent
where, $\rho = \Psi^*\Psi$ and ${\bf J} = ({\bf J}_x, 0, 0)$ are the probability density and the current density respectively, with, ${\bf J}_x = \frac{i\hbar }{36B x}(\Psi\Psi^*_{,x} - \Psi^*\Psi_{,x})$. It is important to note that the continuity equation in the above standard form is found only under the choice of the factor ordering index $n = -1$. Thus, factor ordering index has been fixed from physical argument. Finally, in the very early Universe when $\alpha$ is of the order of Planck's dimension, the term containing the cosmological constant remains subdominant in the effective potential and may be neglected. The extremization of the $V_e$ then yields,

\be a = \sqrt{\frac{k}{2}}(t-t_0),\ee

\noindent
which has been obtained earlier \cite{16}. This clearly depicts that the condition under which the potential is extremum represents coasting solution. Naturally, it is an artefact of curvature squared action. Further, it fixes the curvature parameter to positive value, $k > 0$. In the later epoch, $\Lambda$ term starts playing a dominant role, but it is not possible to obtain a solution of the extremum of the potential in closed form, in the presence of $\Lambda$ term.

\section{Anisotropic metric for a further check}
In the previous section we have shown that in the process of canonical formulation of $R^2$ gravity, the Lapse function acts as Lagrange multiplier, whose variation leads to Hamiltonian constraint equation. But, since only a very special minisuperspace has been accounted for such purpose, it is necessary to check if the same holds in anisotropic minisuperspace too, in order to prove that the result is not an artifact of reducing the theory to a measure zero subset of its configuration space. Therefore, let us take up spatially symmetric Kantowski-Sachs (K-S), Bianchi-I (B-I) and Bianchi-III (B-III) metrics, which can be expressed altogether as,

\be ds^2 = -N(t)^2 dt^2 + a^2 dr^2 + b^2[d\theta^2 + f_k^2 d\phi^2], \ee

\noindent
where, $f_k = sin\theta \Rightarrow k = + 1$ (K-S), $f_k = \theta \Rightarrow k = 0$ (B-I), $f_k = sinh\theta \Rightarrow k = - 1$ (B-III), and for which

\be ^4R = \frac{2}{N^2}\left(\frac{\ddot a}{a}+2\frac{\ddot b}{b}+2\frac{\dot a\dot b}{a b}+\frac{\dot b^2}{b^2} +\frac{k N^2}{b^2}-\frac{\dot a \dot N}{a N}-2\frac{\dot b \dot N}{b N} +\frac{\dot b^2}{b^2}\right).\ee

\noindent
Now, taking $z=a b^2$ and removing total derivative term as before in the action

\be A = \beta\int R^2 \sqrt{-g}\;d^4 x + \sigma_1 + \sigma_2,\ee

\noindent
we have,

\be
A = \int \frac {4 B}{N^3}\Big(\frac{\ddot z^2}{z}-\frac{2 \dot N \dot z \ddot z}{z N}-\frac{4 \dot b \dot z \ddot z}{b z}+\frac{6 \dot b^2 \ddot z}{b^2}+\frac{\dot N^2 \dot z^2}{N^2 z}+\frac{4 \dot b^2 \dot z^2}{b^2 z}-\frac{6 \dot b^2 \dot N \dot z}{b^2 N}-\frac{8 \dot b^3 \dot z}{b^3}+\frac{9 z \dot b^4}{b^4}+\frac{6k N^2 \dot b^2}{b^4}+\frac{k^2 N^4 z}{b^4}\Big)dt+\sigma_2,
\ee

\noindent
where, $\sigma_1$ gets cancelled with the total derivative term $\frac{8 B k \dot z}{N b^2}.$ Let us now introduce auxiliary variable,

\be Q = \frac{\partial A}{\partial \ddot z} = \frac{8B}{N^3} \left[\frac{\ddot z}{z} - \frac{\dot N \dot z}{N z} - 2\frac{\dot b \dot z}{b z}+3\frac{\dot b^2}{b^2}\right]\ee

\noindent
and express the action in the canonical form as,

\be A = \int\left[Q \ddot z - \frac{\dot N\dot z}{N} Q - 2\frac{\dot b\dot z}{b} Q + 3Qz \frac{\dot b^2}{b^2} + 24Bkz \frac{\dot b^2}{N b^4} - \frac{N^3Q^2z}{16B}+\frac{4Bk^2Nz}{b^4}\right]dt + \sigma_2.\ee

\noindent
As before, the first term in the above action (76) is integrated by parts and the total derivative term gets cancelled with $\sigma_2 (= - Q \dot z)$. Finally we are left with,

\be A = \int\left[-\dot Q \dot z - \frac{\dot N\dot z}{N} Q - 2\frac{\dot b\dot z}{b} Q + 3Qz \frac{\dot b^2}{b^2} + 24Bkz \frac{\dot b^2}{N b^4} - \frac{N^3Q^2z}{16B}+\frac{4Bk^2Nz}{b^4}\right]dt.\ee

\noindent
Now, the canonical momenta, $p_Q$ and $p_N$ have the same expressions as found in equation (53) of section 3, while $p_z$ and $p_b$ are different, viz.,

\be p_z = -\dot Q -\frac{\dot N}{N} Q - 2 \frac{\dot b}{b} Q \ee

\be p_b = -2Q \frac{\dot z}{b} -6 \frac{\dot b}{b^2} Q z + 48Bk \frac{z \dot b}{Nb^4}. \ee

\noindent
The $N$ variation equation is given by

\be -\frac{\ddot z Q}{N} - \frac{\dot z\dot Q}{N} + 24Bk\frac{z \dot b^2}{N^2 b^4}+ \frac{3 N^2 Q^2 z}{16B} - 4Bk^2 \frac{z}{b^4} = 0.\ee

\noindent
As in section (3), removing $\ddot z$ term, in view of the definition of the auxiliary variable (75), one can easily verify that this is the Hamiltonian of the system in disguise,

\be H_c = N\left[- \frac{\dot z\dot Q}{N} - \frac{\dot N}{N^2}\dot z Q -2 \frac{\dot b \dot z}{b N} Q + 3 \frac{z \dot b^2}{N b^2} Q + 24Bk\frac{z \dot b^2}{N^2 b^4}+ \frac{N^3 Q^2 z}{16B} - 4Bk^2 \frac{z}{b^4}\right],\ee

\noindent
which is constrained to vanish. Now, using the expression,

\be p_Q p_z = \dot z\dot Q + \frac{\dot N}{N}\dot z Q + 2 \frac{\dot b}{b}\dot z Q,\ee

\noindent
the Hamiltonian constraint equation in terms of the phase space variables is obtained as,

\be H_c = -p_Q p_z - \frac{4 B k^2 N z}{b^4} + \frac{N^3 Q^2 z}{16B} + \frac{b^2N}{12z} \frac{(b p_b - 2Q p_Q)^2}{(N Q b^2 + 8Bk)} = 0.\ee

\noindent
Finally, to express $H_c = N{\mathcal H}$, let us choose as before,

\be x = \frac{\dot z}{N},\ee

\noindent
so that,

\be Q = \frac{\partial A}{\partial \ddot z} = \frac{\partial A}{\partial \dot x}\frac{d\dot x}{d\ddot{z}} = \frac{p_x}{N},\;\;and,\;\;p_{Q} = -\dot z = -N x.\ee

\noindent
Therefore, we need to replace $Q$ by $\frac{p_x}{N}$ and $p_{Q}$ by $-N x$ in the above Hamiltonian. Hence we get,

\be H_c = N\left[x p_z - \frac{4 B k^2 z}{b^4} + \frac{z {p_x}^2}{16B} + \frac{b^2}{12z} \frac{(b p_b + 2x p_x)^2}{(b^2 p_x + 8Bk)}\right]= N{\mathcal{H}} = 0.\ee

\noindent
Thus, here again we observe that the lapse function $N$ acts as Lagrange multiplier. Further, action (76) can now be expressed in the canonical form as

\be A = \int\left(\dot z p_z + \dot x p_x + \dot b p_b - N\mathcal{H}\right)dt~ d^3 x.\ee

\noindent
Note that here to make the calculation simple, we started with $z = ab^2$ which is different from $h_{i j}$. Therefore the above canonical action (87) can not be expressed in terms of the basic variables $(h_{i j}, K_{i j})$ as in equation (65). May be for this reason canonical quantization of the above Hamiltonian (86) is likely to yield a non-hermitian effective Hamiltonian operator, which is not of much interest. A detailed and rigorous calculation in view of Ostrogradski's prescription \cite{14} will be attempted in future.

\section{Einstein-Hilbert action being modified by curvature squared term }

We have shown that the lapse function acts as Lagrange multiplier even in anisotropic minisuperspace metrics. However, since canonical quantization does not yield a hermitian effective Hamiltonian, so we leave anisotropic minisuperspace model and turn our attention to Robertson-Walker minisuperspace model once again. Now, to get a Newtonian analogue, we need to take up Einstein-Hilbert term in addition to $R^2$ gravity. A general coordinate invariant fourth order action is expressed as,

\be A = \int\left[\frac{R - 2\Lambda}{16\pi G} + \beta R^2 +
\alpha C_{\mu\nu\rho\sigma}C^{\mu\nu\rho\sigma}\right]\sqrt{-g}
d^4 x.\ee

\noindent
This action can formally be cast to a unitary renomalizable quantum theory of gravity with positive energy states under Lee-Wick \cite{5} prescription. In the Robertson-Walker minisuperspace metric the Weyl tensor vanishes and after being supplemented by appropriate boundary term, it is expressed as,

\be A = \int\left[\frac{R-2\Lambda}{16\pi G} + \beta R^2\right]\sqrt{-g}d^4x + \sigma+\sigma_1+\sigma_2,\ee

\noindent
where, $\sigma = \frac{1}{8\pi G}\int\sqrt{h} K d^3 x$ is the Gibbons-Hawking-York boundary term. This action leads to inflation without phase transition, followed by reheating \cite{2}. Hence, if one is interested in canonical formulation of a general coordinate invariant fourth order action in Robertson-Walker minisuperspace metric, it is sufficient to start with action (89). It is important to note that action (89) may be cast into a positive definite one under appropriate choice of $\beta$, and has a newtonian gravity long-distance limit as a classical theory. However, a more general quadratic action is expressed as \cite{20},

\be A = \int\left[\frac{R - 2\Lambda}{16\pi G} + \alpha R^2 + \gamma R_{\mu\nu}R^{\mu\nu} + \epsilon R_{\mu\nu\rho\sigma}R^{\mu\nu\rho\sigma}+\lambda \Box R\right]\sqrt{-g} d^4 x.\ee

\noindent
Instead of taking Kretschman scalar squared term, it is customary to express it as $\epsilon\chi$, where,

\[\chi = \frac{1}{32\pi^2}\int\left(R_{\mu\nu\rho\sigma}R^{\mu\nu\rho\sigma}-4 R_{\mu\nu}R^{\mu\nu} + R^2\right)\sqrt{-g} d^4 x.\]

\noindent
The term $\chi$, also known as Gauss-Bonnet term is topologically invariant and so its functional derivative vanishes in four dimension. Further, $\Box R$ is manifestly covariant total divergent term and is usually ignored. Hence, the above action reduces to,

\be A = \int\left[\frac{R - 2\Lambda}{16\pi G} + \alpha R^2 + \gamma R_{\mu\nu}R^{\mu\nu}\right]\sqrt{-g} d^4 x,\ee

\noindent
which is also the action for the gravitational sector under BRST symmetry \cite{21}. The above action may also be recast as,

\be A = \int\left[\frac{R - 2\Lambda}{16\pi G} + \beta R^2 + \gamma(R_{\mu\nu}R^{\mu\nu} - \frac{1}{3}R^2)\right]\sqrt{-g} d^4 x,\ee

\noindent
just by choosing, $\beta = \alpha + \frac{1}{3}\gamma$. In the Robertson-Walker minisuperspace model, $\int(R_{\mu\nu}R^{\mu\nu} - \frac{1}{3}R^2)\sqrt{-g} d^4 x$ is again a total derivative term and thus is ignored. Hence, the above action (92) again reduces to the one given in (89), in the Robertson-Walker metric. It is interesting to note that the Weyl tensor squared term,

\be C_{\mu\nu\rho\sigma}C^{\mu\nu\rho\sigma} = R_{\mu\nu\rho\sigma}R^{\mu\nu\rho\sigma} - 2 R_{\mu\nu}R^{\mu\nu} + \frac{1}{3}R^2,\ee

\noindent
may also be expressed as,

\be C_{\mu\nu\rho\sigma}C^{\mu\nu\rho\sigma} =  (R_{\mu\nu\rho\sigma}R^{\mu\nu\rho\sigma} - 4 R_{\mu\nu}R^{\mu\nu} + R^2) + 2(R_{\mu\nu}R^{\mu\nu} - \frac{1}{3}R^2),\ee

\noindent
where, the first one is the Gauss-Bonnet term. Therefore, under the assumption of homogeneity and isotropy of space time, dubbed as the cosmological principle, both the actions (88) and (90) or (91) in disguise, reduce to (89), which we take up for the present purpose. In fact, instead of expressing (91) in the form given in (92), one could also express it as,

\be A = \int\left[\frac{R - 2\Lambda}{16\pi G} + \beta R_{\mu\nu}R^{\mu\nu} + \gamma(R_{\mu\nu}R^{\mu\nu} - \frac{1}{3}R^2)\right]\sqrt{-g} d^4
x,\ee

\noindent
which reduces to

\be A = \int\left[\frac{R - 2\Lambda}{16\pi G} + \beta R_{\mu\nu}R^{\mu\nu} \right]\sqrt{-g} d^4 x,\ee

\noindent
in the Robertson-Walker minisuperspace model under consideration. This action is more general than (89), since in the gauge theoretic formulation of gravitation, $R^2$ only gives a scalar mode, while $R_{\mu\nu} R^{\mu\nu}$ contains vector mode in the form of massless spin 2 gravitons. Although, in the absence of $R_{\mu\nu}R^{\mu\nu}$ term from the action (91), the spin-two ghost disappears, nevertheless, the divergence problems in the quantum theory reappears \cite{22}. Despite, at present we pay our attention to action (89), since expressing the boundary term as a total derivative term corresponding to $R_{\mu\nu}R^{\mu\nu}$ is still obscure.\\

\noindent
Following the same procedure as in section 2, one can express action (89) in the following canonical form,

\be A = \int\Big[\frac{c}{16\pi G}\Big(-\frac{3\dot{z}^2}{2N\sqrt{z}}+6kN\sqrt{z} -2\Lambda N z^{\frac{3}{2}}\Big) + Q \ddot z - \frac{\dot N\dot z}{N} Q - \frac{N^3 \sqrt z}{36 B} Q^2 + \frac{18Bk\dot z^2}{Nz^{\frac{3}{2}}}+\frac{36BNk^2}{\sqrt z}\Big]dt + \sigma_2.\ee

\noindent
As before, the term $Q\ddot z$ in above action (97) is integrated by parts and the total derivative term gets cancelled with $\sigma_2 ( = - Q \dot z)$, and we are finally left with,

\be A = \int\left[\frac{c}{16\pi G}\left(-\frac{3\dot{z}^2}{2N\sqrt{z}}+6kN\sqrt{z} -2\Lambda N z^{\frac{3}{2}}\right) - \dot Q \dot z - \frac{\dot N\dot z}{N} Q - \frac{N^3 \sqrt z}{36 B} Q^2 + \frac{18Bk\dot z^2}{Nz^{\frac{3}{2}}}+\frac{36BNk^2}{\sqrt z}\right]dt ,\ee

\noindent
where, the auxiliary variable $Q$ and the canonical momenta, $p_Q$ and $p_N$ have the same expressions as found in equations (50) and (53) of section 3, while $p_z$ is different, viz.,

\be p_z = -\frac{3c}{16\pi G}\left(\frac{\dot z}{N\sqrt z}\right) - \dot Q - \frac{\dot N}{N} Q + \frac{36 B k \dot z}{N z^{\frac{3}{2}}}.\ee

\noindent
The $N$ variation equation now is,

\be -\frac{c}{16\pi G}\left(\frac{3\dot{z}^2}{2N^2\sqrt{z}}+6k\sqrt{z} -2\Lambda z^{\frac{3}{2}}\right)-\frac{\ddot z Q}{N} - \frac{\dot z\dot
Q}{N} + 18Bk\frac{\dot z^2}{N^2 z^{\frac{3}{2}}}+ \frac{N^2 \sqrt z}{12B}Q^2  - \frac{36B k^2}{\sqrt z} = 0,\ee

\noindent
which is supposed to be the Hamilton constraint equation. As before, removing $\ddot z$ term, in view of the definition of the auxiliary variable (50), one can easily verify that this is again the Hamiltonian of the system in disguise,

\be H_c = N\left[-\frac{c}{16\pi G}\left(\frac{3\dot{z}^2}{2N^2\sqrt{z}}+6k\sqrt{z} -2\Lambda z^{\frac{3}{2}}\right)-\frac{\dot z \dot Q}{N} - \frac{\dot N\dot z Q}{N^2} + 18Bk\frac{\dot z^2}{N^2 z^{\frac{3}{2}}} + \frac{N^2 \sqrt z}{36B}Q^2  - \frac{36 Bk^2}{\sqrt z} \right]\ee

\noindent
which is constrained to vanish. Now, using the expression,

\be p_Q p_z = \dot z\dot Q + \frac{3c\dot z^2}{16\pi G N \sqrt z} + \frac{\dot N}{N}\dot z Q - \frac{36 B k \dot z^2}{N z^{\frac{3}{2}}},\ee

\noindent
the Hamiltonian constraint equation in terms of the phase space variables is obtained as,

\be H_c = \frac{c}{16\pi G}\left(\frac{3 p_Q^2}{2N\sqrt{z}}-6N k\sqrt{z} +2N\Lambda z^{\frac{3}{2}}\right)-p_Q p_z -  \frac{18 Bk}{N
z^{\frac{3}{2}}}p_{Q}^2 + \frac{N^3 \sqrt z}{36B}Q^2  -\frac{36BN k^2}{\sqrt z} = 0 .\ee

\noindent
Finally, to express $H_c = N{\mathcal H}$, let us choose as before,

\be x = -2K_{ij} = \frac{\dot z}{N}\ee

\noindent
and as a result,

\be Q = \frac{\partial A}{\partial \ddot z} = \frac{\partial A}{\partial \dot x}\frac{d\dot x}{d\ddot{z}} = \frac{p_x}{N},\;\;and,\;\;p_{Q} = -\dot z = -N x.\ee

\noindent
So we need to replace $Q$ by $\frac{p_x}{N}$ and $p_{Q}$ by $-N x$ in the above Hamiltonian. Thus we get,

\be H_c = N\left[x p_z + \frac{\sqrt z}{36B}p_{x}^2 - 18 Bk\left(\frac{x^2}{z^{\frac{3}{2}}}  - \frac{2k}{\sqrt z}\right)+\frac{c}{16\pi G}\left(\frac{3 x^2}{2\sqrt{z}}- 6 k\sqrt{z} +2\Lambda z^{\frac{3}{2}}\right)\right]= N{\mathcal{H}} = 0.\ee

\noindent
It is important to notice that momentum is not associated with the Einstein-Hilbert sector of the action. The action (98) can now be expressed in the canonical form with respect to the basic variables as,

\be A = \int\left(\dot z p_z + \dot x p_x - N\mathcal{H}\right)dt~ d^3 x = \int\left(\dot h_{ij} \pi^{ij} + \dot K_{ij}\Pi^{ij} - N\mathcal{H}\right)dt~ d^3 x,\ee

\noindent
as suggested by Ostrogrdski and at par with expression (65). The corresponding quantum version is

\be \frac{i\hbar}{\sqrt z}\frac{\partial \Psi}{\partial z} = -\frac{\hbar^2}{36B x}\left(\frac{\partial^2}{\partial x^2} + \frac{n}{x}\frac{\partial}{\partial x}\right)\Psi -18 Bk \left(\frac{x}{z^2}+\frac{2k}{z x}\right)\Psi + \frac{c}{16\pi G}\left(\frac{3 x}{2 z}-6 \frac{k}{x} +2\Lambda \frac{z}{x}\right)\Psi.\ee

\noindent
Again under a further change of variable, the above equation takes the look of the Schr{\"o}dinger equation, viz.,

\be i\hbar\frac{\partial \Psi}{\partial \alpha} = -\frac{\hbar^2}{54 B}\left(\frac{1}{x}\frac{\partial^2}{\partial x^2} + \frac{n}{x^2}\frac{\partial}{\partial x}\right)\Psi +\left[\frac{c}{16\pi G}\Big(\frac{x}{\alpha^{\frac{2}{3}}}- \frac{4 k}{x} + \frac{4\Lambda \alpha^{\frac{2}{3}}}{3x}\Big) - 12 B k \Big(\frac{x}{\alpha^{\frac{4}{3}}} +\frac{2k}{x\alpha^{\frac{2}{3}}}\Big)\right]\Psi = \hat H_e\Psi,\ee

\noindent
where, the proper volume $\alpha  = z^{\frac{3}{2}} = a^3$ again plays the role of internal time parameter and the effective potential $V_e$, given by,

\be V_e = \frac{c}{16\pi G}\left(\frac{x}{\alpha^{\frac{2}{3}}} - \frac{4 k}{x} + \frac{4\Lambda \alpha^{\frac{2}{3}}}{3x}\right) - 12 B k \left(\frac{x}{\alpha^{\frac{4}{3}}}+\frac{2k}{x\alpha^{\frac{2}{3}}}\right),\ee

\noindent
is a function of both the so called time variable $\alpha$ and $x$, as before. We observe that Einstein-Hilbert sector appears as effective potential in the quantum version of the theory. We can try to explore the reason for this. In the case of E-H action alone, the momentum is $p_z \propto \dot z$. But here, under the introduction of auxiliary variable, one gets, $p_Q = -\dot z$, while a complicated expression for $p_z$, gives $\dot Q$. With the choice of basic variable, this $p_Q$ has been transformed to a generalized co-ordinate $x$, which reappears in the effective potential. The effective Hamiltonian ($\hat H_e$) is again hermitian and so, one to write the continuity equation for $n = -1$ as before, and the probability interpretation follows as in section 3.

\section{ Classical and semiclassical solution (under WKB approximation)}
\subsection{Classical solution}
Before going into the WKB approximation, let us see if one can generate the classical solution obtained by Starobinsky \cite{1} in view of the field equations corresponding to the action (89). In this connection, let us recall that only for the sake of canonical formulation of the theory, we have introduced an auxiliary variable. Thus it appears that we are having two variables $x$ and $z$. Nevertheless, they are not independent at the classical level, since we are essentially working with fourth order equation having a single variable $a$ - the scale factor. Now the Einstein equations with quantum corrections in one loop approximation is

\be R_{\mu\nu} - \frac{1}{2}g_{\mu\nu} R = 8\pi G <T_{\mu\nu}>.\ee
$<T_{\mu\nu}>$ consists only of local terms which arise in the process of regularization as,

\[ <T_{\mu\nu}> = \frac{k_2}{2880 \pi^2}\left(R_{\mu}^{\lambda} R_{\nu\lambda} - \frac{2}{3}R R_{\mu\nu}-\frac{1}{2}g_{\mu\nu}R_{\lambda\delta} R^{\lambda\delta}+ \frac{1}{4}g_{\mu\nu}R^2\right)\]
\be \hspace{1.3 cm}+ \frac{k_3}{17280\pi^2} \left(2R_{;\mu;\nu} - 2g_{\mu\nu}R^{;\lambda}_{;\lambda}- 2RR_{\mu\nu} + \frac{1}{2}g_{\mu\nu} R^2 \right),\ee
where the constants $k_2$ and $k_3$ are sums of contributions from quantum fields with different spin values. In the isotropic space-time under consideration, the Hamiltonian constraint equation takes the following form

\be \frac{\dot a^2 + k}{a^2} = \frac{1}{{\mathrm{H}}^2}\left(\frac{\dot a^2 + k}{a^2}\right)^2 - \frac{1}{M^2}\left[2\frac{\dot a\stackrel{...}{a}}{a^2}-\frac{\ddot a^2}{a^2}+2\frac{\dot a^2\ddot a}{a^3}-3\frac{\dot a^4}{a^4} - 2k\frac{\dot a^2}{a^4} + \frac{k^2}{a^4}\right],\ee
where ${\mathrm{H}}^2 = \frac{360 \pi}{Gk_2} $ and $M^2 = -\frac{360 \pi}{Gk_3}$. Starobinsky \cite{1} argued ``It is worth
noting that the evolution of the Universe need not follow a `generic' solution, it may well be described just by this unique one, at least initially'' and found that the above field equation (113) is satisfied by the following set of solutions,
\begin{eqnarray}
 a &=& {\mathrm{H}}^{-1}\cosh{{\mathrm{H}} t},\;\;k = +1\\
 a &=& a_0\exp{( {\mathrm{H}}t)},\;\;k=0\\
 a &=& {\mathrm{H}}^{-1}\sinh{{\mathrm{H}} t},\;\;k = -1
\end{eqnarray}
On the contrary any form of the Hamiltonian constraint equations (101), (103) or (106) can be expressed in terms of the scale factor $(a)$ as

\be \frac{\dot a^2 + k}{a^2} = \frac{\Lambda}{3} - \frac{96 B\pi G}{c}\left[2\frac{\dot a\stackrel{...}{a}}{a^2}-\frac{\ddot a^2}{a^2}+2\frac{\dot a^2\ddot a}{a^3}-3\frac{\dot a^4}{a^4} - 2k\frac{\dot a^2}{a^4} + \frac{k^2}{a^4}\right].\ee
The field equations (113) and (117) appear to be different but one can trivially check that the above set of solutions (114) through (116) satisfy the field equation (117) under the condition, ${\mathrm{H}} = \sqrt{\frac{\Lambda}{3}}$.

\subsection{Semiclassical solution under WKB approximation}
Instead of considering the time dependent Schr{\"o}dinger equation (109) let us, for the sake of simplicity, take up the time independent equation (108) for presenting a semiclassical solution in the standard WKB method, expressing it as,

\be -\frac{\hbar^2}{36B}\sqrt z\left(\frac{\partial^2}{\partial x^2} + \frac{n}{x}\frac{\partial}{\partial x}\right)\Psi  - i\hbar x\frac{\partial \Psi}{\partial z} + V\psi = 0 \ee
where
\be V = \frac{c}{16\pi G}\left(\frac{3 x^2}{2\sqrt z}-6 k\sqrt z +2\Lambda z\sqrt z\right) -18 Bk \left(\frac{x^2}{z\sqrt z}+\frac{2k}{\sqrt z}\right).\ee

\noindent
The above equation may be treated as time independent Schr{\"o}dinger equation with two variables $x$ and $z$ and therefore, as usual, let us sought the solution of equation (118) as,
\be \psi = \psi_0e^{\frac{i}{\hbar}S(x,z)}\ee

\noindent
and expand $S$ in power series of $\hbar$ as,

\be S = S_0(x,z) + \hbar S_1(x,z) + \hbar^2S_2(x,z) + .... \ .\ee

\noindent
Now inserting the expressions (120) and (121) in equation (118) and equating the coefficients of different powers of $\hbar$ to zero, one obtains the following set of equations (upto second order)

\begin{eqnarray}
 \frac{\sqrt z}{36B}S_{0,x}^2 + x S_{0,z} + V(x,z)& = & 0 .\\
 -\frac{\sqrt z}{36B}\left[ i S_{0,xx} - 2S_{0,x}S_{1,x} + \frac{i}{x}n S_{0,x} \right] + x S_{1,z} & = & 0.\\
 -\frac{\sqrt z}{36B}\left[ i S_{1,xx} - S_{1,x}^2 - 2S_{0,x}S_{2,x} + \frac{i}{x}n S_{1,x} \right] + x S_{2,z} & = & 0,
\end{eqnarray}
\noindent
which are to be solved successively to find $S_0(x,z),\; S_1(x,z)$ and $S_2(x,z)$ and so on. Now identifying $S_{0,x}$ as $p_x$ and $S_{o,z}$ as $p_z$, one can recover the classical Hamiltonian constraint equation ${\mathcal{H}} = 0$, given in equation (106) from equation (122). Hence $e^{\frac{i}{\hbar}S_0}$ predicts a strong correlation between coordinates and momenta. Therefore, $S_{0}(x, z)$ can now be expressed as,

\be S_0 = \int p_x dx + \int p_z dz, \ee
apart from a constant of integration which may be absorbed in $\psi_0$. The integrals in the above expression can be evaluated using any one of the classical solutions for $k = 0, \pm 1$ presented in equations (114 - 116) and using the definition of $p_z$ given in (99) and $p_x = N Q$, where the expression for $Q$ is given in (75), remembering the relation (105), viz., $N x = \dot z$, where, $z = a^2$. For the present purpose we choose the solution (114) for $k = 1$ and make the standard gauge choice $N = 1$. Note that the condition for the extremum of the effective potential $V_e$ presented in equation (70) requires  $k = +1$. Further we choose $n = -1$, since probability interpretation holds only for such value of $n$.  Using solution (114) $x (= \dot z)$ and the expressions of $p_x$, $p_z$ can be expressed in terms of $z$ as,

\be x = 2\sqrt z\sqrt{\frac{\Lambda}{3}z - 1},\;\;p_x = 36B\left( \frac{2\frac{\Lambda}{3}z - 1}{\sqrt z}\right),\;\;and\;\;p_z = 3\sqrt{\frac{\Lambda}{3}z - 1}\left[ -\frac{c}{16\pi G} - 24B \frac{\Lambda}{3}  + \frac{12B}{z} \right]\ee
and hence the integrals in (125) are evaluated as,

\be \int p_x dx = \int 36B\left( \frac{2\frac{\Lambda}{3}z - 1}{\sqrt z}\right)\left(\frac{2\frac{\Lambda}{3}z - 1}{\sqrt z \sqrt{\frac{\Lambda}{3}z - 1}} \right)dz
= 24B\left[4\left( \frac{\Lambda}{3}z - 1 \right)^{\frac{3}{2}} + 3\arctan\left(\sqrt{\frac{\Lambda}{3}z - 1}\right)\right] \ee

\be \int p_z dz =  -\left( \frac{c}{8\pi G \frac{\Lambda}{3}} + 48B \right)\left( \frac{\Lambda}{3} z - 1 \right)^{\frac{3}{2}} + 72B\left( \sqrt{\frac{\Lambda}{3}z - 1} - \arctan\sqrt{\frac{\Lambda}{3}z - 1} \right).\ee

\noindent
Therefore, explicit form of $S_0$ is,

\be S_0 = \left( -\frac{3 c}{8\pi G \Lambda} + 48B \right)\left( \frac{\Lambda}{3}z - 1 \right)^{\frac{3}{2}} + 72B\sqrt{\frac{\Lambda}{3}z - 1} \ee
and at this end the wave function is

\be \psi = \psi_0 e^{\frac{i}{\hbar}\left[\left( -\frac{3 c}{8\pi G\Lambda} + 48B \right)\left( \frac{\Lambda}{3}z - 1 \right)^{\frac{3}{2}} + 72B\sqrt{\frac{\Lambda}{3}z - 1}\right]}.\ee

\subsection{ First order approximation}
Now equation (123) can be expressed as,

\be i S_{0,xx} - 2S_{0,x}S_{1,x} - \frac{i}{x}S_{0,x} - 72B\frac{2\frac{\Lambda}{3}z - 1}{\sqrt z}S_{1,x} = 0, \ee

\noindent
where we have used the relation $S_{1,z}=S_{1,x}\frac{dx}{dz}$. The above equation may be rearranged as,

\be i\frac{p_{x,x}}{p_x} - i\frac{1}{x} = 2S_{1,x} + 72B\left(\frac{2\frac{\Lambda}{3}z - 1}{\sqrt z}\right)\frac{1}{p_x}S_{1,x} = 4S_{1,x}, \ee

\noindent
which under integration yields the following explicit form of $S_1$, viz.,

\be S_1 = \ln{\left( \frac{p_x}{x} \right)^{\frac{i}{4}}} + f_1(z).\ee
Again rewriting expression (132) in terms of $z$ using (126), one obtains

\be S_{1,z} = \frac{i}{4}\left[\frac{2\frac{\Lambda}{3}}{2\frac{\Lambda}{3}z - 1} - \frac{2\frac{\Lambda}{3}z -1}{2z(\frac{\Lambda}{3}z - 1)} -\frac{1}{2z}\right],\ee
which under integration yields

\be S_1 = \ln\left(\frac{2\frac{\Lambda}{3}z - 1}{z\sqrt{\frac{\Lambda}{3}z - 1}}\right)^{\frac{i}{4}} + f_2(x).\ee
Hence, (133) and (135) together yields,

\be S_1 =  \ln{\left( \frac{p_x}{x} \right)^{\frac{i}{4}}} + \ln\left(\frac{2\frac{\Lambda}{3}z - 1}{z\sqrt{\frac{\Lambda}{3}z - 1}}\right)^{\frac{i}{4}}.\ee
\noindent
So the wave function upto first order approximation reads,
\be \psi = \psi_0 \left( \frac{x}{p_x} \right)^{\frac{1}{4}}\exp^{\frac{i}{\hbar}S_0} = \psi_0 \left( \frac{z^2 (\frac{\Lambda}{3}z - 1)}{18B(2\frac{\Lambda}{3}z-1)^2}\right)^{\frac{1}{4}} e^{\frac{i}{\hbar}\left[\left( -\frac{3 c}{8\pi G\Lambda} + 48B \right)\left( \frac{\Lambda}{3}z - 1 \right)^{\frac{3}{2}} + 72B\sqrt{\frac{\Lambda}{3}z - 1}\right]}\ee

\noindent
Following the above procedure, it is possible in principle to go for the next higher order correction to evaluate $S_2$. However it contributes to the oscillatory part making the form of the wave function complicated. The above form of the wave function is oscillatory, indicating that the region is classically allowed and the wave function is strongly peaked about a set of solutions to the classical field equations. Nevertheless, we have fixed a single classical solution following the argument of Starobinsky \cite{1}, viz., ``It is worth noting that the evolution of the Universe need not follow a `generic' solution, it may well be described just by this unique one, at least initially''. Since the WKB wave function (137) has been evaluated using the inflationary solution (114) for $k = +1$, therefore the above form of the wave function is peaked around the inflationary solution (114).
\section{Summary}

In an earlier work, in the context of isotropic and some anisotropic minisuperspace models (i.e. Bianchi-I, Kantowski-Sachs, Bianchi-III) it was shown that, one can split the boundary term $\Sigma = \int ^4 R K\sqrt{h} d^3 x$, corresponding to $R^2$ action as, $\Sigma = \sigma_1 + \sigma_2$, where both are total derivative terms, under appropriate choice of auxiliary variable \cite{16}. In view of such a boundary term, in section (4) and section (5) it has been shown that the lapse function $N(t)$ acts as Lagrange multiplier both in the isotropic (Robertson-Walker) and anisotropic (Bianchi-I, Kantowski-Sachs and Bianchi-III) minisuperspace models under consideration, as in the case of Einstein-Hilbert action. The action has finally been expressed in the canonical form in terms of the basic variables $h_{ij}, K_{ij}, p_{ij}, \Pi_{ij}$ following Ostrogradski's prescription \cite{14}. For this purpose, following steps are performed.\\
1. The action is written in terms of the basic variables $h_{ij}$, and the existing total derivative term is eliminated upon integration by parts, which gets cancelled with $\sigma_1$.\\
2. Auxiliary variable is introduced, taking derivative of the action with respect to the highest derivative of the field variable present in the action, as suggested by Horowitz \cite{8}.\\
3. The action is then expressed in the canonical form in view  of the said auxiliary variable and upon integration by parts the total derivative term gets cancelled with $\sigma_2$.\\
4. The $N$ variation equation is then identified with the Hamiltonian constraint equation as in the case of Einstein-Hilbert action.\\
5. The canonically conjugate momenta corresponding to the auxiliary variable is replaced by the extrinsic curvature tensor $K_{ij}$ and the auxiliary variable by the momenta canonically conjugate to the extrinsic curvature, following Ostrogradski's prescription \cite{14}. The constrained Hamiltonian can then be cast as $H_c = N{\mathcal H}$, as in the case of Einstein-Hilbert action.\\
6. The action then can be cast in the canonical form $A = \int\left(\dot h_{ij} \pi^{ij} + \dot K_{ij}\Pi^{ij} - N\mathcal{H}\right)dt~ d^3 x$ [equations (65) and (107)], in view of the basic variables ${h_{ij}, K_{ij}, \pi^{ij}, \Pi^{ij}}$ as suggested by Ostrogradski \cite{14}.\\
Canonical quantization is then straightforward, which leads to a Sch\"odinger like equation, where, $\alpha = a^3$, the proper volume, acts as the internal time parameter. The effective Hamiltonian is self adjoint and so the probabilistic interpretation is straightforward again. To express the continuity equation in the standard form, one has to fix up the operator ordering parameter $n = -1$. Since the effective Hamiltonian has time dependence through the effective potential, so the unitary transformation is given by the Dyson series \cite{23}. Hence, there exists a continuous one parameter unitary group $U(\alpha)$, expressed by the Dyson series, Viz.,

\[U(\alpha) = {\mathcal T}\exp{\left[-\frac{i}{\hbar}\int_{\alpha_0}^{\alpha}H_e(x,\alpha)d\alpha\right]},\]

\noindent
where, ${\mathcal T}$ is the Dyson time ordering symbol. Finally, the effective potential is extremum only if the scale factor admits coasting solution for the curvature parameter $k \ge 0$, which implies that inflation is an artefact of curvature squared action for a initially closed Universe. Classical and semi-classical solutions are presented in section 7, the semiclassical wave function is peaked around an inflationary solution. Such a quantum description is totally different from the one that may be obtained using Scalar-Tensor equivalence, since the later produces nothing more than Wheeler-DeWitt equation for a non-minimally coupled scalar field.\\
Although, the whole process of canonization has been carried out in minisuperspace model, nevertheless, it might give insight to extend the work in the whole superspace.

\appendix

\section{Derivation of constrained Hamiltonian under change of variable }

\noindent
In the appendix we show that under suitable choice of the
auxiliary variable, one can get rid of the $N$ variation equation
in the following manner. Let us replace the auxiliary variable $Q$
by $q = NQ$. Thus the action (52) takes the following form,

\[ A = B\int\left[\frac{q\ddot z}{B N} - \frac{q\dot N\dot z}{B
N^2} - \frac{N q^2 \sqrt z}{36 B^2} + \frac{18k\dot
z^2}{Nz^{\frac{3}{2}}}+\frac{36Nk^2}{\sqrt z}- \frac{2N\Lambda
z^{\frac{3}{2}}}{\beta}\right]dt + \sigma_2, \hspace{1.65in} (A.1)\]

\noindent
which after integrating the first term by parts reads,

\[ A = B\int\left[- \frac{\dot q\dot z}{B N} - \frac{N q^2 \sqrt z}{36
B^2} + \frac{18k\dot z^2}{Nz^{\frac{3}{2}}}+\frac{36Nk^2}{\sqrt
z}- \frac{2N\Lambda z^{\frac{3}{2}}}{\beta}\right]dt + \sigma_2 +
\frac{q\dot z}{N}, \hspace{1.7in} ( A.2) \]

\noindent
for which, \\

\[p_q = -\frac{\dot z}{N};\;\;p_z = -\frac{\dot q}{N} +
36B\frac{k\dot z}{N z^{\frac{3}{2}}}  \hspace{4.1in} (A.3)\]

\noindent
and so the $N$ variation equation is

\[ -\frac{B}{N}\left[- \frac{q\dot z}{B N} + \frac{18k\dot
z^2}{Nz^{\frac{3}{2}}} + \frac{N q^2 \sqrt z}{36 B^2}
-\frac{36Nk^2}{\sqrt z} + \frac{2N\Lambda
z^{\frac{3}{2}}}{\beta}\right] = 0, \hspace{2.54in} (A.4) \]

\noindent
while the Hamiltonian is,

\[  H_c = B\left[- \frac{q\dot z}{B N} + \frac{18k\dot
z^2}{Nz^{\frac{3}{2}}} + \frac{N q^2 \sqrt z}{36 B^2}
-\frac{36Nk^2}{\sqrt z} + \frac{2N\Lambda
z^{\frac{3}{2}}}{\beta}\right]. \hspace{2.6in} (A.5)\]

\noindent
Thus the Hamiltonian $H_c$ is constrained to vanish as before. Now
the Hamiltonian can be expressed in terms of the phase space
variable in a straightforward manner as,

\[ H_c = N\left[-p_q p_z - 18Bk\frac{p_q ^2}{z^{\frac{3}{2}}} +
\frac{q^2\sqrt z}{36B}-\frac{36 B k^2}{\sqrt z}+\frac{2B \Lambda
z^{\frac{3}{2}}}{\beta}\right] = N {\mathcal H} = 0. \hspace{1.95in} (A.6) \]

\noindent
To canonically quantize the system, one has to turn over to the
basic variables as mentioned earlier, and this is done by
replacing $q$ by $-p_x$ and $p_q$ by $x$. The Hamiltonian finally
reads,

\[ {\mathcal H} = xp_z + \frac{\sqrt z}{36B}p_x^2  -
18Bk\frac{x^2}{z^{\frac{3}{2}}} - \frac{36B k^2}{\sqrt z}+\frac{2B
\Lambda}{\beta}z^{\frac{3}{2}} = 0,  \hspace{2.82in}   (A.7) \]

\noindent
which is the same one as found in equation (63).

\end{document}